# Are there stars in Bluesky? A comparative exploratory analysis of altmetric mentions between X and Bluesky

Wenceslao Arroyo-Machado[1]; Nicolas Robinson-Garcia[2,*]; Daniel Torres-Salinas[2]
warroyom@asu.edu; elrobin@ugr.es; torressalinas@ugr.es

[1] *Center for Science, Technology and Environmental Policy Studies (CSTEPS), School of Public Affairs, Arizona State University, Phoenix, AZ 85004, USA*
[2] *Department of Information and Communication Sciences, University of Granada, Granada, Spain*

*Corresponding author

**Abstract**

This study examines the shift in the scientific community from X (formerly Twitter) to Bluesky, its impact on scientific communication, and consequently on social metrics (altmetrics). Analyzing 10,174 publications from multidisciplinary and library and information science (LIS) journals in 2024, the results reveal a notable increase in Bluesky activity for multidisciplinary journals in November 2024, likely influenced by political and platform changes, with mentions doubling or quadrupling for journals like Nature and Science. In LIS, the adoption of Bluesky is more limited and shows significant variations across journals, suggesting discipline-specific adoption patterns. However, overall engagement on Bluesky remains significantly lower than on X. While X currently dominates altmetric mentions, the observed growth on Bluesky suggests a potential shift in the future, underscoring its emerging role in academic dissemination and the challenges of adapting scholarly communication metrics across evolving platforms.

ARE THERE STARS IN BLUESKY?

## 1. Introduction

On November 5th, 2024, Donald Trump won the US presidency for a second term over his opponent Kamala Harris, among his closest supporters was Elon Musk, owner of the social media platform X (previously Twitter) after a controversial purchase in October 2022. A day after his triumph, Trump announced that Musk would be head of the Department of Government Efficiency under his administration. Therefore, over the next weeks, millions of users have moved to Bluesky (Holterman, 2024), a rival service also founded by Jack Dorsey, founder of Twitter. Bluesky basically mimics the functionalities that Twitter had before Musk's take over. Among the users shifting, many of them seem to be researchers, pushing Altmetric to announce on December 3rd, that they will now be also tracking this new social media platform (Altmetric, 2024).

User migration across social media platforms is a common aspect of social media and affects the platforms where scientists share their results and interact with their peers (Jeong et al., 2023). Consequently, it also impacts the metric aspects of scholarly information, influencing how altmetrics are collected (e.g., Google Plus or Delicious). However, this is an interesting case due to two reasons. First, the difficulty to track social media discussion of publications has increased since Musk's takeover, pointing towards a strategy from Altmetric to redirect the traffic of mentions to scientific literature from X to Bluesky. Second, if that is the case, this move has important implications, as X, contrary to other deceased platforms, is the source representing the largest bulk of data offered by Altmetric.

The first objective of this brief communication is to verify whether there is evidence of a community shift from X to the new platform, Bluesky, from its inception to November. Consequently, if such migration is confirmed, the second objective is to examine the differences in values and indicators between the two platforms. This study analyzes the coverage and number of mentions in two distinct sets of journals: three multidisciplinary journals and four library and information science journals in 2024, comparing their presence on both X and Bluesky. This analysis will help us understand whether the activity reflected on Bluesky is comparable in volume to that on X, a crucial step in assessing the implications of incorporating Bluesky mentions and evaluating their significance or comparability to those from X. The results are expected to provide valuable insights into whether Bluesky currently serves as an effective platform for the dissemination of scientific information and the generation of altmetric indicators.

## 2. Methodology

We constructed a dataset of 10,174 publications from three multidisciplinary journals, — Nature, Science, and Proceedings of the National Academy of Sciences of the United States of America—, and four Information Science & Library Science (LIS) journals—Journal of the Association for Information Science and Technology (JASIST), Journal of Informetrics (JoI), Scientometrics, and Quantitative Science Studies (QSS)— for the year 2024. This selection was made to ensure representation of both broad, multidisciplinary research and specialized fields, providing two distinct control groups to justify the robustness of our comparisons. We focus solely on this year as Bluesky opened the registration to its service in February 2024, before one could open an account by invitation only.

We extracted records' DOI, resolved DOI URL, title and alternative URLs associated with the record, such as OA version or PDF, as provided by OpenAlex. Mentions from Bluesky were extracted by querying its API for each of the links. In the case of X, we queried the Altmetric API for records' DOIs as Altmetric already solves the issue of multiple links. Hence, we would add to our dataset the total number of mentions provided by each source. The Python script used for data retrieval from Bluesky and X via Altmetric is available in GitHub (https://github.com/Wences91/bluesky_altmetrics). The data exploration was performed using descriptive statistics in R.

## 3. Results

Table 1 shows the coverage and average number of accounts mentioning each paper for both platforms. In the case of Science, Nature and PNAS, we observe that, except for the latter, the levels of coverage are relatively similar, slightly lower in Bluesky. To gain a more accurate understanding, it is essential to consider the number of accounts mentioning scientific publications, where it becomes evident that the figures are notably lower for Bluesky. For instance, in Nature, the average number of accounts on X is 148.49 compared to only 4.86 on Bluesky. Similarly, for Science, the average is 103.68 on X and 9.42 on Bluesky. Focusing on Library and Information Science (LIS) journals, *Scientometrics*, *Journal of Informetrics*, and *QSS* manage to have approximately 50% of their articles mentioned on X. In contrast to multidisciplinary journals, this indicator decreases significantly on Bluesky. For example, *Scientometrics* achieves only 15%, while *JASIST* and *QSS* reach 25%. Furthermore, the drop in the number of accounts mentioning articles is particularly notable on Bluesky. For instance, *Scientometrics* decreases from an average of 17.21 accounts on X to just 2.09 on Bluesky, while *QSS* experiences an even sharper decline, from 28.58 to 3.57.

**Table 1.** *Coverage and average number of accounts mentioning papers by journal across X and Bluesky*

| Journal | Nr Papers | X | | Bluesky | |
|---|---|---|---|---|---|
| | | Papers mentioned | Avg. Accounts | Papers mentioned | Avg. Accounts |
| *Nature* | 4089 | 4059 (99%) | 148.49 | 3742 (92%) | 4.86 |
| *PNAS* | 3613 | 3010 (83%) | 27.91 | 1159 (32%) | 6.08 |
| *Science* | 1924 | 1437 (75%) | 103.68 | 1360 (71%) | 9.42 |
| *Scientometrics* | 305 | 143 (47%) | 17.21 | 47 (15%) | 2.09 |
| *Journal of Informetrics* | 107 | 28 (26%) | 5.04 | 7 (6%) | 1.57 |
| *JASIST* | 80 | 42 (53%) | 13.60 | 19 (24%) | 2.11 |
| *QSS* | 56 | 26 (46%) | 28.58 | 14 (25%) | 3.57 |
| **Total** | **10,174** | **8745 (86%)** | **96.01** | **6348 (62%)** | **6.03** |

ARE THERE STARS IN BLUESKY?

Given that user migration occurred in November, we analyzed the chronological evolution of mentions on Bluesky (Figure 1). A significant increase in mentions was observed during November for the three multidisciplinary journals. Notably, Nature and Science both experienced substantial growths, with mentions doubling and quadrupling, respectively, reaching over 4,000 mentions in November. Similarly, PNAS, although with fewer overall mentions, quadrupled its figures, reaching 600 mentions. In the field of Library and Information Science (LIS), where mentions are generally limited, only QSS followed a similar pattern, with mentions increasing in November to 25. The remaining LIS journals displayed an irregular pattern over time, without any notable increases during the final month analyzed.

**Figure 1.** *Evolution of Bluesky mentions of scientific publications from four major altmetrics journals between December 2023 and November 2024*

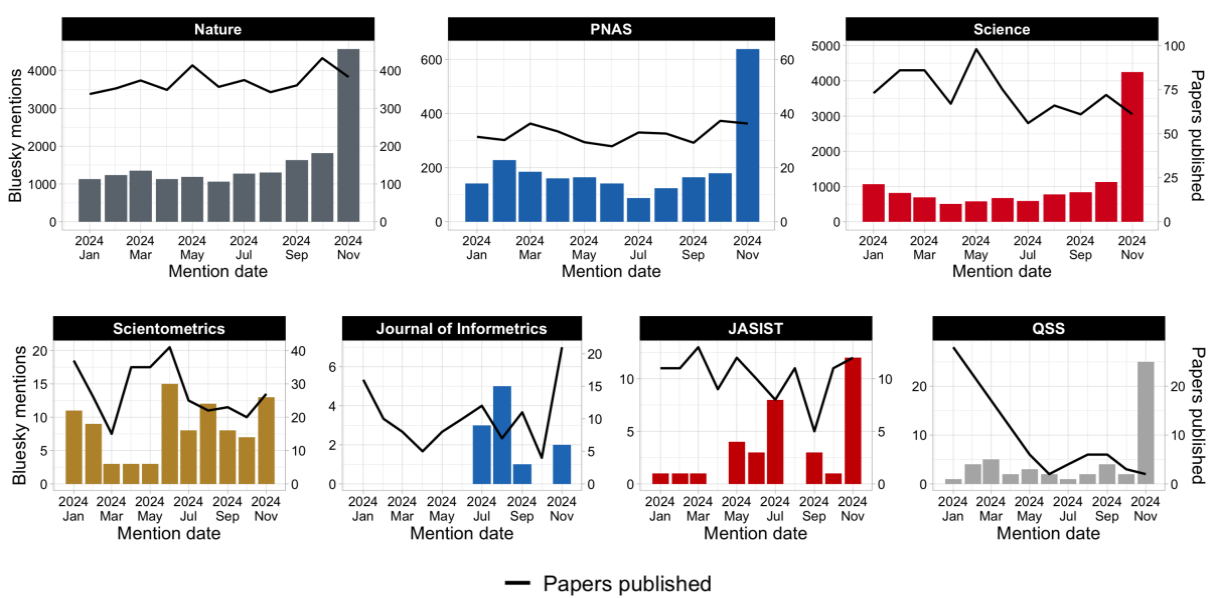

Figure 2 illustrates the percentage distribution of users mentioning November 2024 papers on Bluesky and X across different journals, highlighting a clear dominance of X for five journals, with mentions ranging from 88% to 97%. Bluesky, in contrast, accounts for a much smaller share of mentions, generally under 12% for multidisciplinary journals such as Nature, Science, and PNAS. Notable exceptions are observed in the Library and Information Science (LIS) category: in Journal of Informetrics, mentions are evenly distributed between Bluesky (50%) and X (50%), while in JASIST, Bluesky surpasses X with 67% of mentions compared to 33%. These findings suggest that while X remains the primary platform for most disciplines, Bluesky shows signs in LIS journals, indicating potential differences adoption in platform engagement across fields.

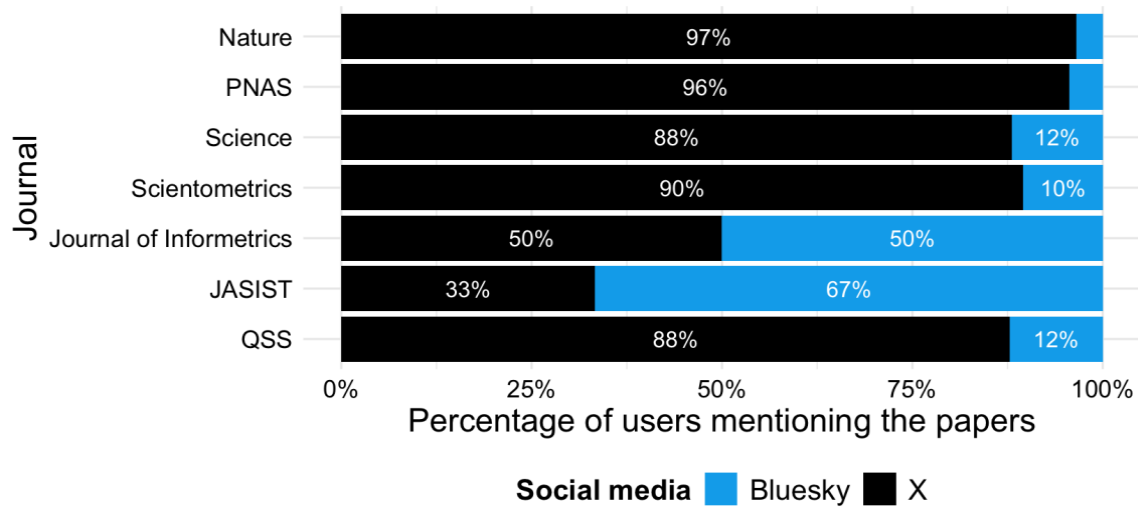

**Figure 2.** *Percentage of users mentioning November 2024 papers on Bluesky and X*

## 4. Discussion

Our exploratory analysis shows some evidence of user migration between platforms that could justify the inclusion of Bluesky by Altmetric, especially when looking at the top journals which are normally the ones attracting the highest share of mentions (REF). However, this is not clear when focusing on LIS journals, pointing at different levels of adoption by fields. This could question to what extent this migration across platforms has taken place by fields. Furthermore, when looking at the distribution of accounts per platform, we observe that it is JASIST, which is a US journal, the one with the highest share of accounts coming from Bluesky, reflecting to some extent how politics get intertwined with academia.

In this sense, user migration across platforms is a slow process which does not necessarily mean an abandonment of one of the platforms, favouring the other (Hou & Shiau, 2019), but tends to follow a push and pull model which could well end up with both platforms coexisting. This potential scenario could lead to further questions as to the already intricate question on the meaning of altmetric mentions (Robinson-Garcia et al., 2017), as we have two social media platforms which seem to be similar in functionalities but can potentially be hosting very distinct communities of users. Going back to the question on the moment in which Altmetric has decided to index Bluesky, it is certainly surprising as nothing of the sort happened with previous hypes such as the user migration to Mastodon two years ago (Chan, 2022). We can only speculate as to the reasons, one being the increasing opacity of X. However, this may need for re-investigation on the forming of communities discussing scientific literature (Arroyo-Machado et al., 2021) and adding complexity as there may be complementarities between both platforms.

In conclusion, while evidence of platform migration exists, activity on Bluesky remains limited, particularly regarding LIS journals, which show low levels of mentions and account activity. Future scenarios could see this trend increase and grow further or stabilize at current levels. Regardless, X continues to dominate in terms of altmetric activity, suggesting it remains the primary platform for scholarly engagement.